# Entangled States and Entropy Remnants of a Photon-Electron System

**S Varró**

Research Institute for Solid State Physics and Optics, H-1525 Budapest, POBox 49, Hungary

E-mail: varro@mail.kfki.hu

**Abstract**. In the present paper an example of entanglement between two different kinds of interacting particles, photons and electrons is analysed. The initial-value problem of the Schrödinger equation is solved non-perturbatively for the system of a free electron interacting with a quantized mode of the electromagnetic radiation. Wave packets of the dressed states so obtained are constructed in order to describe the spatio-temporal separation of the subsystems before and after the interaction. The joint probability amplitudes are calculated for the detection of the electron at some space-time location and the detection of a definite number of photons. The analytical study of the time evolution of entanglement between the initially separated electron wave packet and the radiation mode leads to the conclusion that in general there are non-vanishing entropy remnants in the subsystems after the interaction. On the basis of the simple model to be presented here, the calculated values of the entropy remnants crucially depend on the character of the assumed switching – on and – off of the interaction.

**Keywords:** Entanglement, Strong-field photon-electron interactions, Quantum mechanical wave packets.

**PACS:** 03.65.Ud, 42.50.Hz, 03.65.-w, 42.50.Dv

## 1. Introduction

The concept of entanglement has long been introduced by Schrödinger (1935), who was motivated by the criticism expressed by Einstein *et al* (1935) on the conceptual foundation of quantum mechanics, which that time drew attention to some seemingly paradoxical features associated to the new theory. Concerning this kind of fundamental questions we refer the reader, for example, to the book by Bohm (1951) and the collection of the excellent papers by Bell (2004). More than thirty years after the first reliable experiments by Wu and Shaknov (1950), Aspect *et al* (1982) measured correlations between entangled photon pairs, proving a strong violation of Bell's inequalities, in agreement with the predictions of quantum theory. In the meantime it turned out that entanglement plays a crucial role in the nowadays rapidly developing branches of science, namely in quantum information theory (see e.g. Alber et al 2001, Bouwmeester et al 2001 and Stenholm and Suominen 2005) and in quantum communication and quantum computing (see e.g. Bendjaballah *et al* 1990, Williams 1999 and Nielsen and Chuang 2000).

Besides entanglement between particles of the same kind with some discrete degrees of freedom, there has recently been a growing interest in the study of continuous-variable entanglement between different kinds of particles. For instance, Fedorov and coworkers in a series of papers (see Fedorov *et al* 2004, 2005, 2006 and 2007) have thoroughly analysed from this point of view the wave packet dynamics in breakup processes, like ionization of atoms and dissociation of molecules. To some extent, the subject of the present paper belongs to the cathegory of processes, in that we shall deal with the dynamics of free electron wave packets accompanied by photon exchange. The photon-electron interaction in free space is the most fundamental phenomenon of quantum electrodynamics (QED), and the nature of the very high-order multiphoton processes is still the subject of an extensive research nowadays. In the following we shall study entanglement in high-order photon emission and absorption processes at the simplest imaginable level, namely, we stay in the framework of non-relativistic quantum mechanics, where the interaction is modelled by the minimal coupling term in dipole approximation. In case of a *free electron* the single-mode description of the interacting radiation may be justified by that we shall take a highly occupied mode (which, in turn may also be considered as a representative of an assembly of modes in a narrow spectral range). It should also be stressed that the discussion on the spatial separation of the two subsystems before and after the interaction will rely to a large extent on introducing a phenomenological switching –on and –off  of the interaction.

Besides quantum optics, the subject of the present paper belongs also to the 'physics of strong-field phenomena'. Perhaps it not superfluous to quote some basic references in this field, at least concerning the mathematical description of high-order interactions of the quantized radiation and an electron. The first reference to be quoted is undoubtely the paper by Bloch and Nordsieck (1937), in which they have given the first *non-perturbative analytic* treatment of the interaction of a free electron with the *whole assembly* of the quantized radiation field (in the 'soft-photon limit'), and in this way solved the problem of infrared divergences appearing in *perturbative* QED (see also Bogoliubov and Shirkov (1959)). The *exact* solution of the Dirac equation of the system consisting of an electron and a *single quantized plane wave mode* of the radiation field has first been presented by Bersons (1969), which ment that he constructed a generalization of the famous Volkov (1935) states, which are *exact* solutions of the Dirac equation of an electron in the presence of an *external plane wave* of the radiation field (of arbitrary high intensity). This states have played a very important role in the theory of strong-field interactions from the beginnings (just to mention some early basic works, see e.g. by Brown and Kibble on the nonlinear Compton scattering, Bunkin and Fedorov (1965) on optical tunneling, Ritus and Nikishov (1979) on pair creation in a strong laser field). Concerning strong-field physics today in general, we refer the reader to the recent review papers by Mourou *et al* (2006) on relativistic optics, Ehlotzky *et al* (2009) on relativistic scattering in strong laser fields and Krausz and Ivanov (2009) on attosecond physics.

The important point for us here is that, however initial and final dressed states of the free electron embedded in the quantized radiation field have been used so far, these dressed states have exclusively been taken as product states of an electron momentum eigenstate and the state of the photon field. In the frame of such a description one cannot consider entanglement. One immediately encounters entangled photon-electron states if one constructs wave packets of the dressed states in order to describe more realistically the spatio-temporal dynamics of the interaction process. In a recent paper (Varró (2008), henceforth referred to as I) we have constructed Gaussian wave packets (with respect to the electron's momentum) of the stationary states of the system of an electron and a quantized mode of the electromagnetic radiation. On the basis of this exact analytic treatment, the essential characteristics of the system (e.g. photon number distribution, von Neumann entropy and linear entropy) were determined. However this states are not solutions of a true initial value problem, rather they are merely joint stationary states,leaving the question unanswered, how could this states be generated? The present work is devoted to the discussion of this problem in the simplest framework.

In Section 2 we present the exact solutions of the Schrödinger equation of the interacting photon-electron system developing from a product state parametrized by the electron's initial momentum and by the initial occupation number of the quantized mode. In Section 3 we shall construct Gaussian wave packets of these solutions, which are normalized entangled states developing from an uncorrelated product state of an electron wave packet and a number state. In Section 4 the reduced density operators shall be derived for these states, with the help of which and analysis of the time evolution of the entropies shall be carried out. In Section 5 a brief summary of the results and conclusions closes our paper.

## 2. Solution of the initial value problem of the interacting photon-electron system

Let us consider the joint system consisting of a quantized mode of the radiation field and a Schrödinger electron. For the sake of simplicity, we take a circularly polarized plane wave for the mode, because in this case the $\vec{A}^2$ term of the Hamiltonian is diagonal in the photon number state basis. Moreover, we shall use dipole approximation, and consider electronic motions in the $x - y$ plane (the electron's motion along the $z$ -direction is a simple free propagation). The Hamiltonian of the system reads

$$H = \left[ \frac{1}{2m} \left( \hat{\vec{p}} + \frac{e}{c} \vec{A} \right)^2 + \hbar\omega(A^+ A + 1/2) \right] = \frac{\hat{p}^2}{2m} + \hbar\Omega(A^+ A + 1/2) + \frac{ea}{mc} \hat{\vec{p}} \cdot (\vec{\varepsilon} A + \vec{\varepsilon}^* A^+) , \qquad (1)$$

$$\vec{A} = a(\vec{\varepsilon} A + \vec{\varepsilon}^* A^+), \ a \equiv (2\pi\hbar c^2 / \omega L^3)^{1/2}, \ \Omega \equiv \omega(1 + \omega_p^2 / 2\omega^2), \ \omega_p^2 = 4\pi e^2 / mL^3. \qquad (2)$$

In the above equations $\vec{\varepsilon} = (\vec{\varepsilon}_x + i\vec{\varepsilon}_y) / \sqrt{2}$  is the complex polarization vector (for right circular polarization, when the field is assumed to propagate in the negative $z$ -direction), $\omega$ is the circular frequency of the mode, and $L^3$ is the quantization volume. The photon creation and annihilation operators are denoted by $A^+$ and $A$, respectively, and $AA^+ - A^+A = 1_{photon}$, where the right hand side is the unit operator of the Hilbert space of the photon states. The parameters $-e$, $m$ and $c$ are the electron's charge, mass, the velocity of light in vacuum, respectively, and $\hbar$ denotes Planck's constant divided by $2\pi$. Notice that $\omega_p$ is formally nothing else but the plasma frequency of an electron gas of density $1/L^3$. Even in the case $L \sim \lambda = 2\pi c / \omega$, the ratio $\omega_p / \omega$ is much smaller than unity for usual electron beams and for optical frequencies, however in case of a terahertz radiation it may not be neglected. In obtaining the right hand side of Eq. (1) we have taken into account that $\vec{\varepsilon}^* \cdot \vec{\varepsilon} = 1$ and $\vec{\varepsilon}^* \cdot \vec{\varepsilon}^* = \vec{\varepsilon} \cdot \vec{\varepsilon} = 0$. The solutions to the Schrödinger equation $i\hbar \partial_t |\Psi(t)\rangle = H |\Psi(t)\rangle$, describing the time evolution of the system with the above Hamiltonian, has been presented by Bergou and Varró (1981a). On the basis of this result, the explicit form of the solutions can be brought to the form

$$|\Psi(t)\rangle = \exp\left[ -\frac{i}{\hbar} \frac{\hat{p}^2}{2m(t)} t - i\Omega(A^+A + 1/2)t \right] \cdot D[\hat{\sigma}(\vec{p}, t)] \cdot |\Psi(0)\rangle, \tag{3}$$

where

$$D[\hat{\sigma}(\vec{p}, t)] \equiv \exp[\hat{\sigma}^*(\vec{p}, t)A - \hat{\sigma}(\vec{p}, t)A^+], \qquad m(t) \equiv \frac{1 + \omega_p^2 / 2\omega^2}{1 + (\omega_p^2 / 2\omega^2)[2(\sin\Omega t)/(\Omega t) - 1]} m, \tag{4}$$

and

$$\hat{\sigma}(\vec{p}, t) \equiv -(\vec{\hat{p}} \cdot \vec{\varepsilon}^*) \frac{ea}{mc\hbar\Omega}(e^{i\Omega t} - 1). \tag{5}$$

Since in realistic cases $\omega_p / \omega << 1$, the dressed mass $m(t)$ is practically identical with the bare mass $m$. Anyway, it is interesting to note that, in principle, this effective mass $m(t)$ can be much larger than the bare mass (if $\omega$ approaches $\omega_p / \sqrt{2}$ from above), moreover, it can also be negative (if $\omega < \omega_p / \sqrt{2}$), which corresponds negative energies or, formally, imaginary momenta. As a result, instead of the usual spreading of an electronic wave packet, a contraction would take place. Similar phenomena can certainly be more realizable e.g. in case of the scattering of resonant atoms off the vacuum field at the entrance to a microwave cavity, as was discussed by Scully and Sargent (1972). In the present paper we shall not discuss the question of the possible physical relevance of the formal appearance of the negative mass, and henceforth (after Eq. (6)) we shall denote $m(t)$ and $\Omega$ simply by $m$ and $\omega$, respectively. As is well-known, the displacement operators of the kind displayed in Eq. (4) have an important role in the quantum theory of optical coherence and coherent states, as was first shown by Glauber (1963).

The solution given by Eq. (3) looks particularly simple if the initial state is a product of a momentum eigenstate and a generalized coherent state of the quantized mode, i.e. if $|\Psi(0)\rangle = |\vec{p}\rangle \otimes D[\alpha] n_0\rangle$, then

$$|\Psi(t \mid \vec{p}, n_0, \alpha)\rangle = |\vec{p}\rangle e^{-(i/\hbar)[p^2/2m(t)]t} \otimes D[\sigma(\vec{p}, t)e^{-i\Omega t}] \cdot D[\alpha e^{-i\Omega t}] \cdot |n_0\rangle e^{-i(n_0 + 1/2)\Omega t}, \tag{6}$$

where $\alpha$ is proportional with the initial complex amplitude of the expectation value of the electric field strength of the mode. Because of the property $D(\sigma)D(\alpha) = D(\sigma + \alpha) \exp[i \operatorname{Im}(\sigma\alpha^*)]$ of the displacement operators, has the structure $\sim |\vec{p}\rangle \otimes D(\beta) |n_0\rangle$, so it is still a product state for $t > 0$, too. If $\alpha = 0$ and $n_0 = 0$, then the initial vacuum state $|0\rangle$ goes over to the ordinary coherent state $|\sigma\rangle$, thus, one may say that (at least, according to the present very simplified description) the self radiation field of the free electron is in a coherent state. We also note that, owing to the unitarity of the displacement operators, the exact solutions given by Eq. (6) form a complete orthogonal set on the product Hilbert space $H_{photon} \otimes H_{electron}$ for any parameter $\alpha$, i.e.

$$\langle \Psi(t \mid \vec{p}', n', \alpha) | \Psi(t \mid \vec{p}'', n'', \alpha) \rangle = \delta_2(\vec{p}' - \vec{p}'')\delta_{n', n''}, \qquad (\forall \alpha) \tag{7a}$$

$$\int d^2 p \sum_{n=0}^{\infty} \left| \Psi(t \mid \vec{p}, n, \alpha) \right\rangle \left\langle \Psi(t \mid \vec{p}, n, \alpha) \right| = 1_{photon} \otimes 1_{electron}. \qquad (\forall \alpha) \qquad (7b)$$

The photon statistics of the generalized coherent state of the type $D(\sigma)|n\rangle$, is determined by the matrix elements

$$c_{k,n} \equiv \langle k | D(\sigma) | n \rangle = \begin{cases} (n!/k!)^{1/2} \sigma^{k-n} L_n^{k-n}(|\sigma|^2) e^{-|\sigma|^2/2}, & (k \geq n) \\ (k!/n!)^{1/2} (-\sigma^*)^{n-k} L_k^{n-k}(|\sigma|^2) e^{-|\sigma|^2/2}, & (0 \leq k < n) \end{cases}, \qquad (8)$$

where $L_n^s$ denote generalized Laguerre polynamials (see e.g. Gradshteyn and Ryzhik 2000, formula 8.970.1). To our knowledge, the matrix elements of the type given by Eq. (8), was first obtained by Bloch and Nordsieck (1937), who applied them in the case of $n = 0$, i.e. all modes were considered initially in vacuum state, and then the nowadays well-known coherent states with the associated Poisson photon number distributions $|\sigma|^{2k} \exp(-|\sigma|^2)/k!$ resulted. In the opposite case of large excitations ($n \gg 1$) these matrix elements can be brought to a more tractable form

$$\langle n_0 + k | D[\sigma(\vec{p},t)e^{-i\omega t}] | n_0 \rangle = e^{-ik(\omega t - \chi - \eta)} J_k \left( 2\sqrt{n_0} \mid \sigma(\vec{p},t) \mid \right) + O(n_0^{-3/4})$$
$$= e^{-ik(\omega t - \chi - \eta)} J_k \left( \sqrt{2} \frac{eA_0 \mid h(t) \mid}{mc\hbar\omega} p \right) + O(n_0^{-3/4}) \qquad , \qquad (9)$$

where we have used an asymptotic formula of Hilb's type (see e.g. Erdélyi 1953, formula 10.15(2), or Eqs. (A14) and (A15) in I). In Eq. (9) $J_k(z)$ denotes ordinary Bessel functions of first kind of order $k$, and we have introduced the azimuth angle $\chi$ of the momentum vector by the definition $\vec{p} = p(\cos\chi, \sin\chi)$. We have also introduce the quantity $A_0 = (c/\omega)\sqrt{2\pi\rho\hbar\omega}$, by taking the definition of $\sigma$ in Eq. (5) into account. In fact, $A_0$ is an equivalent to the amplitude of the classical vector potential $\vec{A}_{cl} = A_0(\vec{\varepsilon} e^{-i\omega t} + \vec{\varepsilon}^* e^{i\omega t})$ associated with the photon density $\rho = n_0 / L^3$, if we make the identification $u = E_{cl}^2 / 4\pi = \rho\hbar\omega$. Here $u$ denotes the energy density of the mode, with $\vec{E}_{cl} = -\partial \vec{A}_{cl} / \partial ct = F_0(\vec{\varepsilon}_x \sin\omega t - \vec{\varepsilon}_y \cos\omega t)$ being the electric field strength with amplitude $F_0 = (\omega/c)A_0\sqrt{2} = \sqrt{4\pi\rho\hbar\omega}$ (which may contain a slow time-dependence). In Eq. (9) the function $|h(t)|$ represents a slow time-dependence of the modulus of $\sigma(\vec{p},t)$. More precisely, $h(t)$ is defined by the equation

$$h(t) \equiv |h(t)| e^{i\eta(t)} \equiv \omega^2 \int_0^t dt' \int_0^{t'} dt'' f(t'') e^{i\omega(t-t'')}, \qquad (10)$$

where $f(t)$ is a dimensionless switching function of the electric field, i.e. $E_0(t) = F_0 f(t)$. In obtaining Eq. (10) we have taken into account the classical relation $\vec{E} = -\partial \vec{A} / \partial ct$.

In fact, the present analysis to follow can also be applied for the consideration of the more general case when the electric field strength has a form $\vec{E} = F_0 f(t + z/c)[\vec{\varepsilon}_x \sin\omega(t + z/c) + \vec{\varepsilon}_y \cos\omega(t + z/c)]$, where $f$ is an envelope function. For motions of the electron restricted to the $z = 0$ plane the envelope function may be considered as modelling the switching-on and -off of the interaction. Of course, with the use of a phenomenological switching function, one has to keep in mind that this procedure, according to the usual Fourier description, implicitely assumes excitations of an assembly of modes in some spectral range $\Delta\nu$ around the central frequency.

At the end of the present section we note that the matrix elements given by Eq. (9) can be derived from the semiclassical Schrödinger equation of the electron interacting with the external field $\vec{E}_{cl}(t)$ given above. Really, if we take as solution a momentum eigenstate, then this contains a factor with periodic phase modulation, and this factor can be decomposed into the Fourier series, by using the Jacobi-Anger formula (see e.g. Erdélyi 1953, formula 7.2.4 (26), or Eq. (A4) in I),

$$\exp\{-i[\mu(t)p/\hbar k]\sin(\omega t - \chi - \eta)\} = \sum_{k=-\infty}^{\infty} J_k[\mu(t)pc/\hbar\omega]e^{-ik(\omega t - \chi - \eta)},  \tag{11}$$

where

$$\mu(t) \equiv \mu_0 \mid h(t) \mid, \qquad \mu_0 \equiv eA_0\sqrt{2}/mc^2 = eF_0/mc\omega = 10^{-9}\sqrt{I}/E_{ph}, \qquad F_0 \equiv \sqrt{4\pi(n_0/L^3)\hbar\omega} \tag{12}$$

In Eq. (12) we have defined the "dimensionless intensity parameter" $\mu_0$, which playes an important role in strong-field physics. Its numerical value can be express in terms of the peak intensity $I$ of the radiation field measured in W/cm$^2$, and of the photon energy $E_{ph}$ measured in eV. According to the above discussion, we have also introduced the amplitude of the electric field strength $F_0 \equiv (\omega/c)A_0\sqrt{2} = \sqrt{4\pi\rho\hbar\omega}$, where the photon density defined as $\rho = n_0/L^3$ (kept fixed as both $n_0$ and $L$ going to infinity). In this way, the Fourier coefficients in Eq. (11) coincide, with an accuracy of order $n_0^{-3/4}$, with the matrix elements given by Eq. (9). Consequently, in the large-intensity limit we are allowed to use the 'quasi-classical' formula

$$\langle n_0 + k \mid D[\sigma(\vec{p},t)e^{-i\omega t}] \mid n_0 \rangle \rightarrow J_k\left(2\sqrt{n_0} \mid \sigma(\vec{p},t) \mid\right)e^{-ik(\omega t - \chi - \eta)} = J_k[\mu(t)pc/\hbar\omega]e^{-ik(\omega t - \chi - \eta)}, \tag{13}$$

where the notations are the same as in Eqs. (12) and (9).

## 3. Entangled photon-electron states in the case of high initial photon excitations

The entangled photon-electron states developing from a pure initial state being the product of a number state and an electron wave packet is taken in the form

$$\left| \Psi_g(t) \right\rangle \equiv \int d^2p\, g(\vec{p} - \vec{p}_0)e^{-(i/\hbar)\vec{p}\cdot\vec{r}_0} \left| \Psi(t \mid \vec{p}, n_0, 0) \right\rangle, \quad g(\vec{p}) \equiv g(p) = (w/\hbar\sqrt{\pi})\exp(-p^2w^2/2\hbar^2), \tag{14}$$

where $g$ has been specialized to a Gaussian weight function of width $w$, and $\left| \Psi(t \mid \vec{p}, n_0, 0) \right\rangle$ is a special case of the state given by Eq. (6) with $\alpha = 0$. Owing to the orthonormality property displayed in Eq. (7a), these time-dependent packets are normalized, too. The physical situation to which the state given by Eq. (14) may be associated is the following. At time $t = 0$ the electron is located sharply around the central position $\vec{r}_0$ with an initial drift momentum $\vec{p}_0$, and it is injected into the interaction region (or exposed the radiation) which is swept by a light pulse propagating along the $z$-direction. The longitudinal motion of the electron is assumed to be very slow, and it is also assumed that it stays close to the $z = 0$ plane. Since the coupling of the $z$-motion is negligible in this case, we consider simply a planar motion in the $x-y$ plane.

In order to have an explicit form of the state defined by Eqs. (14) and (6), we express it in the electron's coordinate representation, and, at the same time, expand it in terms of the photon number eigenstates. Thus we write

$$\left| \Psi_g(t) \right\rangle = \int d^2r \left| \Xi(\vec{r},t) \right\rangle \left| \vec{r} \right\rangle = \sum_{k=-n_0}^{\infty} \left| \Phi_k(t) \right\rangle \left| n_0 + k \right\rangle, \tag{15}$$

where

$$\left| \Xi(\vec{r},t) \right\rangle \equiv \sum_{k=-n_0}^{\infty} \Psi_k(\vec{r},t) \left| n_0 + k \right\rangle, \qquad \left| \Phi_k(t) \right\rangle = \int d^2r\, \Psi_k(\vec{r},t) \left| \vec{r} \right\rangle. \tag{16}$$

It is clear that the states $\left| \Xi(\vec{r},t) \right\rangle \in H_{photon}$ and $\left| \Phi_k(t) \right\rangle \in H_{electron}$ satisfy the normalization conditions

$$\int d^2r \left\langle \Xi(\vec{r},t) \mid \Xi(\vec{r},t) \right\rangle = 1, \qquad \sum_{k=-n_0}^{\infty} \left\langle \Phi_k(t) \mid \Phi_k(t) \right\rangle = 1, \tag{17}$$

where the scalar products, of course, are meant in the Hilbert spaces $H_{photon}$ and $H_{electron}$, respectively. The summation index in the above equations has been shifted merely for later convenience. According to Eqs. (15) and (16), the wave packet solution $\left| \Psi(t) \right\rangle$ can be expressed as a joint sum and an integral,

$$\left| \Psi_g(t) \right\rangle = \sum_{k=-n_0}^{\infty} \int d^2r\, \Psi_k(\vec{r},t) \left| \vec{r} \right\rangle \left| n_0 + k \right\rangle, \qquad \Psi_k(\vec{r},t) \equiv \left\langle n_0 + k \mid \left\langle \vec{r} \mid \Psi_g(t) \right\rangle, \tag{18}$$

thus, this solution describes mixed continuous and discrete entanglement.

In order to characterize the entangled photon-electron states $\left|\Psi_g(t)\right\rangle$, defined in Eqs. (14) and (6), we have to determine the expansion coefficients expressed by scalar products in the second equation of Eq. (18). Having shifted the momentum variable by $\vec{p}_0$, the integral becomes

$$\Psi_k(\vec{r},t) \equiv \left\langle n_0 + k \left|\left\langle \vec{r} \right| \Psi_g(t) \right\rangle\right. = e^{i\phi} \int_0^\infty dp\, pg(p)(1/2\pi\hbar) \exp\{-i[p^2 t/2m\hbar]\}$$
$$\times \int_0^{2\pi} d\chi \exp\left[(i/\hbar) pr(t)\cos(\chi - \varphi(t))\right]\left\langle n_0 + k \left| D\left[\sigma(\vec{p}_0 + \vec{p}, t)e^{-i\omega t}\right] \right| n_0 \right\rangle \quad , \quad (19)$$

where $\phi \equiv \vec{p}_0 \cdot (\vec{r} - \vec{r}_0)/\hbar - p_0^2 t/2m\hbar - (n_0 + 1/2)\omega t$, and we have introduced the polar decomposition of the shifted position by the relation $r(t)(\cos\varphi(t), \sin\varphi(t)) \equiv \vec{r} - \vec{r}_0 - (\vec{p}_0/m)t$. The displacement operator in Eq. (19) can be factorized into the product of the displacement operators (the additional phase factor goes to zero in the quasi-classical limit) containing only $\vec{p}_0$ and $\vec{p}$. Then, by inserting the unit operator of the Hilbert space of the mode in the suitable form, applying the general formula given by Eq. (13), and by using the same integration method as in Appendix A in I, we obtain

$$\Psi_k(\vec{r},t) = \frac{e^{i\phi}}{w\sqrt{\pi}} \frac{i^k e^{-ik[\omega t + \varphi(t) - \eta]}}{(1 + it/\tau)} \exp\left[-\frac{(\mu(t)\lambdabar/w)^2 + (r(t)/w)^2}{2(1 + it/\tau)}\right] ,$$
$$\times \sum_{l=-\infty}^{\infty} I_{k-l}\left[\frac{(\mu(t)\lambdabar/w)\cdot(r(t)/w)}{(1 + it/\tau)}\right] J_l[\mu(t)p_0 c/\hbar\omega]i^{-l}e^{il[\chi_0 - \varphi(t)]} \quad , \quad (20)$$

where

$$1/\tau \equiv \hbar/mw^2, \quad \mu(t) \equiv \mu_0 \mid h(t)\mid, \quad r(t)(\cos\varphi(t), \sin\varphi(t)) \equiv \vec{r} - \vec{r}_0 - (\vec{p}_0/m)t, \quad \lambdabar \equiv \lambda/2\pi = c/\omega. \quad (21)$$

In Eq. (20) $I_n(z)$ denotes a modified Bessel functions of the first kind of order $n$ (see Section 7.2 in Erdélyi (1953)), and the polar representation $p_0(\cos\chi_0, \sin\chi_0) \equiv \vec{p}_0$ has been introduced. The characteristic time $\tau$ of the spreading of the wave packet strongly depends on the initial width $w$ is well-known. According to the definition given by Eq. (19), the space-time functions $\Psi_k(\vec{r},t)$ have a clear physical meaning, namely, they are joint probabilities of the simultaneous events when the electron is at position $\vec{r}$ and at the same time $k$ 'excess photons' are excited or deexcited around the large mean photon number $n_0$. At the start of the interaction at $t = 0$ the coupling is zero, i.e. $\mu(0) = 0$, and from Eq. (20) we see that these functions reduce the simple form $\Psi_k(\vec{r},0) = \delta_{k,0}e^{i\phi}\psi_e(\vec{r},t)$, where $\psi_e$ is a freely evolving electron wave packet. This means that we recover the the initial state $\left|\Psi(0)\right\rangle$ in accord with the definition in Eq. (18). The expansion coefficients given by Eq. (20) are time-dependent generalizations of the stationaty ones, for presented in Eq. (22) in I. Here we allow non-vanishing $\vec{p}_0 \neq 0$ and $\vec{r}_0 \neq 0$, which is needed to describe *localized propagation* into the interaction region and the *separation* in space and time from that region. We may say that above we have derived a mathematical background, on the basis of which we can take over the classical intuition in modelling the scattering process. From Eq. (20) we see that the coupling between the photons and the electron are governed by the intensity parameter $\mu_0 = 10^{-9} I^{1/2}/E_{ph}$, which has been already defined in Eq. (12). The (presumably slowly varying) envelope function $h(t)$ (for the definition see Eq. (10)) describes the details of the switching–on and –off of the interaction between the electron and the radiation field. I can be shown that the normalization conditions in Eq. (17) are satisfied if the expansion coefficients are that in Eq. (20), just obtained. This means that the proper normalization 'survives' the quasi-classical limit.

In the special case when $\vec{p}_0 = 0$, $\mid h(t)\mid = 1$ and $\varphi = 0$ the photon states $\left|\Xi(\vec{r},t)\right\rangle$ determined with the expansion coefficients in Eq. (20) have a close connection with a class of number-phase minimum uncertainty states. The states of the mode which minimize the uncertainty product of the photon number and

the Susskind and Glogower (1964) cosine operator have been obtained by Jackiw (1968). Concerning this point we refer the reader to sections 2 and 4 of I (in particular see Eq. (27) of I).

At the end of the present section we would like to note that the coupling constant $\mu$ has very transparent physical meanings. If one solves the Newton equations $m\ddot{x} = -eE_x$ and $m\ddot{y} = -eE_y$ of the electron in the presence of the external electric field $\vec{E} = F_0(\vec{\varepsilon}_x \sin \omega t + \vec{\varepsilon}_y \cos \omega t)$, then it is easily seen that $\mu$ is just the ratio of the amplitude oscillation to the wave length of the radiation. If this ratio approaches unity, then the dipole approximation is not applicable in the case of a free electron. On the other hand, since $\mu$ is the ratio of the velocity amplitude of oscillation to the velocity of light, it is clear that the nonrelativistic description is not justified, either, if $\mu$ gets close (or above) unity. The dimensionless quantity $\mu\lambda/w$ in the expression of the joint probability $\Psi_k(\vec{r},t)$ in Eq. (20) is the ratio of the amplitude of classical oscillation of the electron to the initial transverse width of the wave packet. The classical oscillations mentioned above could have been derived and incorporated into the wave packet dynamics, if we would have used the '$\vec{r}\cdot\vec{E}$ - gauge' for the interaction, but from the point of view of entanglement this is not needed here. Concerning the choice of gauges in quantum optics we refer the reader to the excellent books by Loudon (2000), Scully and Zubairy (1997) and Schleich (2001) from the extensive literature.

## 4. Time evolution of the entropies of the photon-electron system and entropy remnants after the interaction

In the present section we shall determine the reduced density operator of the quantized mode of the radiation field interacting with the electron, and follow the dynamics of the interaction by assuming two kinds of switching.

In order to obtain the reduced density operator $\hat{P}$ of the quantized mode associated to the entangled state $\left|\Psi_g(t)\right\rangle$, we use the expansion coefficients $\Psi_k(\vec{r},t)$ given by Eq. (20), and the electron's position representation to calculate the trace. The partial trace (denoted by $Tr'$) of the dyad $\left|\Psi_g(t)\right\rangle\left\langle\Psi_g(t)\right|$ can be expressed as

$$\hat{P} \equiv Tr'\left\{\left|\Psi_g(t)\right\rangle\left\langle\Psi_g(t)\right|\right\} = \sum_{k=-\infty}^{\infty}\sum_{l=-n_0}^{\infty} P_{kl}\left|n_0+k\right\rangle\left\langle n_0+l\right|, \qquad P_{kl} \equiv \int d^2r\,\Psi_k(\vec{r},t)\Psi_l^*(\vec{r},t). \tag{22}$$

The integral on the right hand side of the second equation in Eq. (22) can be evaluated analytically, yielding

$$P_{kl} = e^{-i(k-l)\alpha}e^{-q}\sum_{n=-\infty}^{\infty} I_{k-n}(q)J_n(\kappa)J_{n-(k-l)}(\kappa), \qquad q \equiv \frac{1}{2}\left[\mu(t)(\lambda/2\pi w)\right]^2, \qquad \kappa \equiv \frac{\mu(t)cp_0}{\hbar\omega}, \tag{23}$$

where we have introduced the phase angle $\alpha \equiv \omega t - \eta - \chi_0$. Since we are investigating the quasi-classical limiting case ($n_0 \to \infty$), henceforth the lower limit of the sum will be taken $-\infty$. With the help of the sum rules $\sum_{k=-\infty}^{\infty} I_k(z) = e^z$ and $\sum_{n=-\infty}^{\infty} J_n^2(z) = 1$, it can be easily proved that $Tr\hat{P} = 1$, so our quasi-classical analytic results obtained are consistent with the necessary requirement for probabilities. By taking the time average $\overline{P}_{kl}$ of $P_{kl}$ over one cycle, only the diagonal terms survive, thus from Eq. (23) we receive the result

$$\overline{P}_{kl} = \delta_{kl}p_k, \qquad p_k \equiv \sum_{n=-\infty}^{\infty} I_{k-n}(q)e^{-q}J_n^2(\kappa), \qquad \sum_{k=-\infty}^{\infty} p_k = 1. \tag{24}$$

For small values of either of the arguments, from the general expression in Eq. (24), for the probability distribution of the photon occupation number can be immediately derived,

$$p_k = I_k(q)e^{-q} \quad (\kappa \ll 1), \qquad \text{and} \qquad p_k = J_k^2(\kappa) \quad (q \ll 1). \tag{25}$$

We note that each limiting expressions in Eq. (25) satisfy the proper normalization condition. If $\vec{p}_0 = 0$, then $\kappa$ is exactly zero, and the first equation of Eq. (25) contains no approximation. For simplicity, in the numerical examples we have used the first approximate formula in Eq. (25). On the basis of the diagonal expression in Eq. (24) we are able to write down immediately an explicit for the von Neumann entropy of the photon field

$$S_{photon}[\hat{P}] \equiv -Tr[\hat{P}\log\hat{P}] \rightarrow -Tr[\overline{P}\log\overline{P}] = S_{photon}[\{p_k\}] \equiv -\sum_{k=-\infty}^{\infty} p_k \log p_k, \qquad (26)$$

which can be calculated numerically. On the other hand, by now we have not been able to diagonalize the electron's density operator (which can also be calculated analytically). A possibility to get around this difficulty, and quantify in a simple way the entanglement, the calculation of the of the linear entropy $H$ offers itself. The linear entropy has been used by several authors (see e.g. Zurek et al 1993 and Joos et al 2003), because it is much easier to calculate, since the diagonalization of the density operator is not needed. The definition of $H$ and its explicite form for the distribution in Eq. (24) read

$$H \equiv 1 - Tr(\overline{P}^2), \quad K \equiv 1/Tr(\overline{P}^2), \quad H_{electron} = H_{photon}[\overline{P}] = H_{photon}[\{p_k\}] \equiv 1 - \sum_{k=-\infty}^{\infty} p_k^2,$$

$$H = 1 - e^{-2q} \sum_{n,m=-\infty}^{\infty} I_{n-m}(2q) J_n^2(\kappa) J_m^2(\kappa). \qquad (27)$$

In deriving the last equation we have used the summation formula $\sum_{l=-\infty}^{\infty} I_{n+l}(z) I_l(z) = I_n(2z)$ for the modified Bessel function. The Schmidt number $K = 1/(1 - H) \geq 1$ is closely related to the linear entropy, and in some cases it is better to visualize the degree of bipartite entanglement with the dependence of this quantity on the input parameters. In the case when $\kappa \ll 1$, from Eq. (27) we obtain

$$H = 1 - I_0(2q)e^{-2q} \qquad\qquad [\ p_k = I_k(q)e^{-q} \qquad (\kappa \ll 1)\ ], \qquad (28)$$

and, on the other hand, in the case when $q \ll 1$, we have from Eq. (27)

$$H = 1 - \sum_{k=-\infty}^{\infty} J_k^4(\kappa) \qquad\qquad [\ p_k = J_k^2(\kappa) \qquad (q \ll 1)\ ]. \qquad (29)$$

In the numerical examples we have used the following two enevelope functions for the electric field strength of the highly populated photon mode

$$f_1(t) = \sin[(\pi/T_1)(t - t_0)], \quad f_2(t) = \sin^2[(\pi/T_1)(t - t_0)] \qquad (t_0 \leq t \leq t_0 + T_1). \qquad (30)$$

In each cases the interaction is limited to the interval $t_0 \leq t \leq t_0 + T_1$, out of which $f_1$ and $f_2$ are zero, such that continuous at the points of the switching –on and –off. The corresponding 'interaction functions' $h_1(t)$ and $h_2(t)$, according to the definition in Eq. (10) can be determined by elementary calculations.

The results of the present section are numerically illustrated in some special cases in the following figures. In Fig. 1 the time-dependence of the envelope functions, (a): $f_1(t)$ and (c): $f_2(t)$ defined in Eq. (30), and in (b): $|h_1(t)|$ and (d): $|h_2(t)|$ the time-dependence of the moduli of the 'interaction function' defined in Eq. (10) are shown. Though both $f_1(t)$ and $f_2(t)$ are continouos, according to the qualitative behaviour of the corresponding $h$-functions, $f_2(t)$ is 'smooth' in comparison with $f_1(t)$. At the end of the interaction (in the present case at the time $t = 31$) neither of them vanishes. We have proved that this is true for any rational ratio $T_1/T$. It is interesting to note that $\mathrm{Re}[h_2(t)]$ is necessarily zero at $T_1$ if $T_1/T$ is an integer.

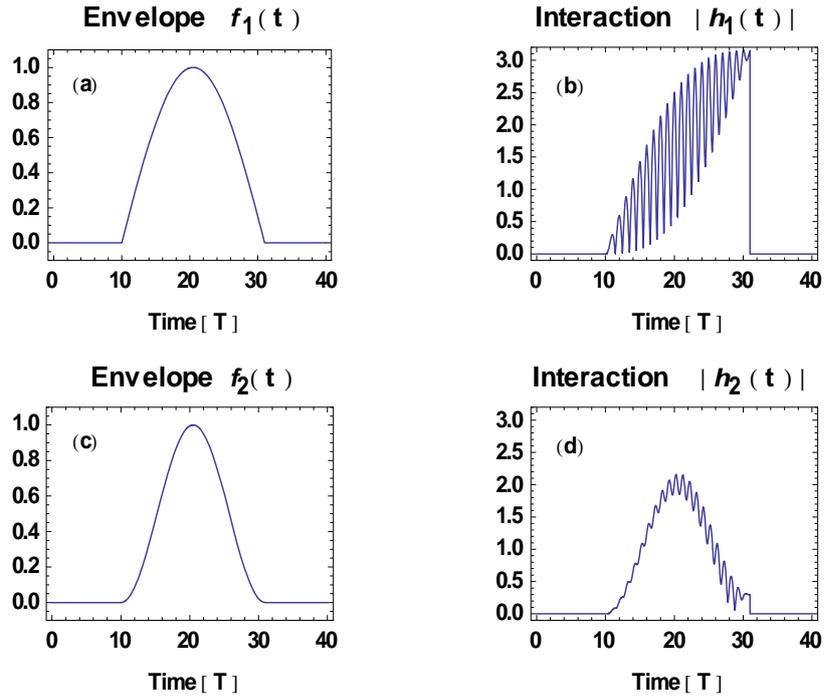

**Figure 1.** Shows the time-dependence of the envelope functions, (a): $f_1(t)$ and (c): $f_2(t)$ defined in Eq. (30), where the unit of time is the period $T$ of the radiation mode. In each figure the pulse is switched – on at $t = 10 \times T$, and switched – off at $t = 31 \times T$, that is the duration of the interaction is $T_1 = 21 \times T$. Figures (b): $|h_1(t)|$ and (d): $|h_2(t)|$ show the time-dependence of the moduli of the 'interaction function' defined in Eq. (10), by specializing the envelope functions to $f_1(t)$ and $f_2(t)$, respectively. On average $|h_1(t)|$ increases with time, on the other hand $|h_2(t)|$ roughly follows the shape of its switching function.

In Fig. 2 we see the time-evolution of the von Neumann entropies (a): $S_1$, (c): $S_2$ and the Schmidt numbers (b): $K_1$, (d): $K_2$ of the photon number distribution given by Eqs. (26), and (27), respectively, in the special case when $\vec{p}_0 = 0$, and for the numerical value $q_0 = 2$. As is seen in these figures, in each cases of the two different switching functions there are 'entropy remnants' left in the subsystems after the interaction was swithed – off. The mathematical reason for that neither of the functions $h_1(T_1) \neq 0$ and $h_2(T_1) \neq 0$ gets exactly to zero at the time of termination of the interaction, as clearly illustrated in Fig. 1. Moreover, the root of these entropy remnants is the still entangled state operator $|\Psi(T_1)\rangle\langle\Psi(T_1)|$ of the complete system, which, due to the absence of the intraction evolves freely in later times $t > T_1$, according to the unpertubed free Hamiltonian. The entanglement survives to some extent in each cases because the final density operator of the complete state (at $t = T_1$) is *not* factorized to a simple product of the form $P_{photon} \otimes P_{electron}$. This can be immediately seen from the functional form of the joint expansion coefficients $\Psi_k(\vec{r}, t = T_1)$.

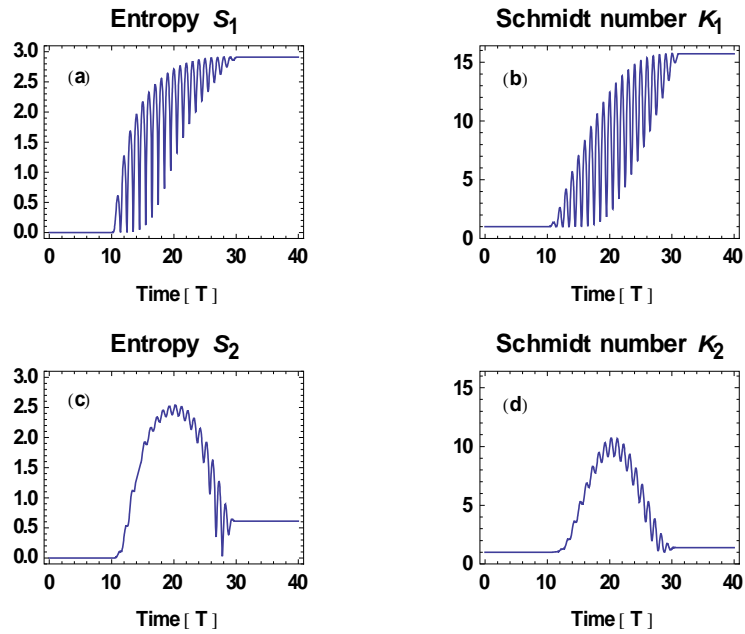

**Figure 2.** Shows the time-evolution of the von Neumann entropies (a): $S_1$, (c): $S_2$ and the Schmidt numbers (b): $K_1$, (d): $K_2$ of the photon number distribution given by Eqs. (26), and (27), respectively, in the special case when $\vec{p}_0 = 0$, and for the numerical value $q_0 = 2$. The subscripts refer to the switching function used in the calculation. In the defining equation of $q$, Eq. (23), we have set $\mu(0) = \mu_0 = 10^{-3}$ and $\lambda / 4\pi w = 10^3$. This means that for the optical radiation we have taken the wavelength $\lambda = 10^{-4} \, cm$ and the intensity $I = 10^{12} \, W / cm^2$. For the initial width of the electron wave packet $w \approx 10^{-8} \, cm$, i.e. one Ångstöm has been assumed. The time scales in this figure are the same as that in Fig. 1. As is seen in these figures, in each cases of the two different switching functions there are 'entropy remnants' left in the subsystems after the interaction was swithed – off.

## 5. Summary

In the present paper we have discussed interactions between photons and electrons, and derived exact analytic expressions for the entangled state of the system evolving from an initial product state representing an electronic wave packet and a number eigenstate of the quantized photon mode. The von Neumann entropy of the photon, and the Schmidt number of the photon and of the electron have been presented in the quasi-classical limit. Since we have solved the initial value problem exactly, we were able to study the time evolution of the entropies of these simple subsystems, and draw some conclusions concerning the question of reversibility and irreversibility of the interaction.

We have made a comparison between two cases distinguished by the different envelope functions modelling the switching –on and –off of the interaction between a light pulse and a localized electron. Though each of these functions are continuous at the instants of switchings, the time evolutions of the system in the two cases are qualitatively different. In one case there is an accumulation of the entropy by the end of the interaction, in the other case the system is almost recovering to a pure state, i.e. it makes an almost reversible cycle. However there are always some entropy remnants present at the end of the process, and the photon-electron system gets off the interaction region in an entangled state, though they are already separating from each other. On the basis of our analytic results we have presented a few numerical illustrations of the existence of the mentioned entropy remnants.

Finally we note that there are more or less straightforward ways towards the generalization of our analysis beyond the dipole approximation and towards the relativistic description.

**Acknowledgements.** This work has been supported by the Hungarian National Scientific Research Foundation OTKA, Grant No. K73728.

**Caption to Figure 1**
Shows the time-dependence of the envelope functions, (a): $f_1(t)$ and (c): $f_2(t)$ defined in Eq. (30), where the unit of time is the period $T$ of the radiation mode. In each figure the pulse is switched – on at $t = 10 \times T$, and switched – off at $t = 31 \times T$, that is the duration of the interaction is $T_1 = 21 \times T$. Figures (b): $|h_1(t)|$ and (d): $|h_2(t)|$ show the time-dependence of the moduli of the 'interaction function' defined in Eq. (10), by specializing the envelope functions to $f_1(t)$ and $f_2(t)$, respectively. On average $|h_1(t)|$ increases with time, on the other hand $|h_2(t)|$ roughly follows the shape of its switching function.

**Caption to Figure 2**
Shows the time-evolution of the von Neumann entropies (a): $S_1$, (c): $S_2$ and the Schmidt numbers (b): $K_1$, (d): $K_2$ of the photon number distribution given by Eqs. (26), and (27), respectively, in the special case when $\vec{p}_0 = 0$, and for the numerical value $q_0 = 2$. The subscripts refer to the switching function used in the calculation. In the defining equation of $q$, Eq. (23), we have set $\mu(0) = \mu_0 = 10^{-3}$ and $\lambda / 4\pi w = 10^3$. This means that for the optical radiation we have taken the wavelength $\lambda = 10^{-4} cm$ and the intensity $I = 10^{12} W / cm^2$. For the initial width of the electron wave packet $w \approx 10^{-8} cm$, i.e. one Ångstöm has been assumed. The time scales in this figure are the same as that in Fig. 1. As is seen in these figures, in each cases of the two different switching functions there are 'entropy remnants' left in the subsystems after the interaction was swithed – off.